\documentclass[a4paper,11pt]{article}
\usepackage[scale={0.75,0.9},centering,includeheadfoot]{geometry}

\usepackage{articlestyle}
\usepackage{booktabs}
\usepackage{mathtools}

\graphicspath{{figs/}}
\usepackage{tikz}

\usepackage{marginnote} 

   \usepackage{algorithm}
   \usepackage{algpseudocode}
  \algdef{SE}[DOWHILE]{Do}{doWhile}{\algorithmicdo}[1]{\algorithmicwhile\ #1}%
\makeatother
\title{\bf{Spatially adaptive covariance tapering}}
\author{\scshape{David Bolin and Jonas Wallin}\\{\small Mathematical Sciences, Chalmers and University of Gothenburg}}
\date{}
\title{Spatially adaptive covariance tapering}
\begin{document}

\maketitle
\begin{center}\begin{minipage}{0.9\textwidth}
\noindent{\bf Abstract:} Covariance tapering is a popular approach for reducing the computational cost of spatial prediction and parameter estimation for Gaussian process models. However, tapering can have poor performance when the process is sampled at spatially irregular locations or when non-stationary covariance models are used. This work introduces an adaptive tapering method in order to improve the performance of tapering in these problematic cases. This is achieved by introducing a computationally convenient class of compactly supported non-stationary covariance functions, combined with a new method for choosing spatially varying taper ranges. Numerical experiments are used to show that the performance of both kriging prediction and parameter estimation can be improved by allowing for spatially varying taper ranges. However, although adaptive tapering outperforms regular tapering, simply dividing the data into blocks and ignoring the dependence between the blocks is often a better method for parameter estimation.

\vspace{0.3cm}\noindent{\bf Key words:}
Kriging; Sparse matrices; Compactly supported covariances; non-stationary covariances; Maximum likelihood
\end{minipage}\end{center}

\section{Introduction}
Gaussian processes are important for statistical analysis of spatial data. In applications the goal is often to predict the process at unobserved locations, which is done by computing the conditional mean of the field given the data. This is referred to as the kriging prediction in geostatistics, and requires solving $\mv{\Sigma}^{-1}\mv{x}$ where $\mv{x}$ is a vector with observed values of the field and $\mv{\Sigma}$ is the covariance matrix for the field at the observation locations. Thus, if the field is observed at $N$ locations, the computational cost for kriging prediction is in general $\mathcal{O}(N^3)$. This limits the applicability for large datasets and is commonly referred to as the ``big N'' problem.

Covariance tapering is a popular method for handling the big N problem. The basic idea of this method is to set small elements in $\mv{\Sigma}$ to zero, which enables the use of computationally efficient sparse matrix techniques for computing $\mv{\Sigma}^{-1}\mv{x}$. For spatial problems, this typically reduces the computational complexity from $\mathcal{O}(N^3)$ to $\mathcal{O}(N^{3/2})$. The simplest way to introduce zeros in $\mv{\Sigma}$ is to replace the covariance function, $r(\mv{h})$, of the Gaussian process with some compactly supported covariance function $T(\mv{h})$, such as a Wendland function \citep{wendland95,gneiting02}. An alternative is to replace $r(\mv{h})$ with a tapered version $r_{tap}(\mv{h}) = r(\mv{h})T(\mv{h})$. The covariance range of $T(\mv{h})$, referred to as the taper range,  determines the sparsity of the resulting covariance matrix.

Several authors have studied the effect of tapering in the case when $r(\mv{h})$ is a Mat\'ern covariance function \citep{matern60}. \cite{kaufman2008} showed that certain parameters of the Mat\'ern covariance function can be consistently estimated using tapering, and \cite{du2009}, \cite{shaby2012}, and \cite{wang2011fixed} further studied the asymptotic properties of tapered estimators in the case of Mat\'ern covariance models. \cite{furrer06} showed that tapering can be used with asymptotically negligible loss in the case of kriging prediction for Mat\'ern fields, and \cite{stein2013taper} extended some of these results beyond the Mat\'ern model.

However, asymptotic results do not guarantee that the method is practically useful. \cite{stein2013taper} used numerical experiments to show that covariance tapering often does not work as well for parameter estimation as simply splitting the observations into blocks and ignoring the dependence between the blocks. Furthermore, \cite{bolin13comparison} showed that tapering also can perform poorly for kriging prediction compared with other methods for handling the big N problem. A reason for this is that the computational cost for stationary tapering depends on the ratio of the taper range and the average spacing of the observations, whereas the accuracy depends on the ratio of the taper range and the range of $r(\mv{h})$. The accuracy also depends on how the observations are located spatially, and more sparsely observed areas will have higher errors for tapered kriging predictions. This often leads to higher errors close to the boundaries of the observation domain as well as ``spotty'' predicted surfaces where sparsely observed regions are biased towards zero.

\cite{stein2013taper} proposed using a simple deformation method to improve performance near the boundaries of the domain. However, this method only works for rectangular domains and does not help in sparsely observed regions. \cite{anderes2013} proposed a more sophisticated deformation method based on quasi-conformal maps where an adaptive taper $T(\mv{s},\mv{t}) = T(\|\varphi(\mv{s}) - \varphi(\mv{t})\|)$ was defined using a stationary taper $T$ and a warping function $\varphi$. A drawback with this approach is that estimation of the warping function $\varphi$ is highly computationally demanding, which limits the usefulness of the method.

This works introduces a new adaptive taper method that is more flexible than that by \cite{stein2013taper} while at the same time being more computationally efficient than the method by \cite{anderes2013}. This is achieved by first introducing a computationally convenient class of compactly supported non-stationary covariance functions in Section \ref{sec:tapers}. How to use this for kriging prediction is then discussed in Section \ref{sec:kriging}. The section first introduces a new method for choosing spatially varying taper ranges and then investigates the accuracy of the method using simulated data. Section \ref{sec:estimation} presents a method for using the adaptive tapers for parameter estimation and investigates the accuracy of the method using numerical experiments. The conclusion of these comparisons is that adaptive tapering outperforms regular tapering, but simply dividing the data into blocks and ignoring the dependence between the blocks is often a better method for parameter estimation. This is in line with what \cite{stein2013taper} found.  Finally, Section \ref{sec:conclusions} contains comments and suggestions for future research.

\section{Adaptive covariance tapers}\label{sec:tapers}
A popular method for constructing non-stationary covariance models is to use the process convolution approach \citep{barry96,higdon01,cressie02,rodriges10}, where a Gaussian stochastic field, $X(\mv{s})$, is defined as the
convolution of a Brownian sheet $\mathcal{B}$ and some convolution kernel $k_{\mv{s}}(\mv{u})$. The covariance function of $X(\mv{s})$ is then given by the integral $C(\mv{s},\mv{t}) = \int k_{\mv{s}}(\mv{u})k_{\mv{t}}(\mv{u}) \md\mv{u}$, which is non-stationary if the kernel changes spatially. One can also note that the support of the covariance function is determined by the support of the kernel. The idea here is therefore to use compactly supported and non-stationary kernels in the process-convolution approach to construct adaptive tapering functions. Thus, we choose the tapering function $T(\mv{s},\mv{t})$ as
\begin{equation}\label{eq:kerneltaper}
T(\mv{s}, \mv{t}) = \int k_\mv{s}(\mv{u}) k_\mv{t}(\mv{u}) d\mv{u}.
\end{equation}

The advantage with this approach is that $T(\mv{s},\mv{t})$ by construction is a valid covariance function for any square integrable $k_\mv{s}(\mv{u})$. The disadvantage, however, is that numerical methods often are required to evaluate the integral in \eqref{eq:kerneltaper}. This is computationally expensive and therefore reduces usefulness of the construction. A crucial property of the tapers we construct below is that the integral can be solved analytically, which means that one has an explicit form of the non-stationary taper function.

\subsection{Hyperspherical tapers}\label{sec:Hyper_taper}
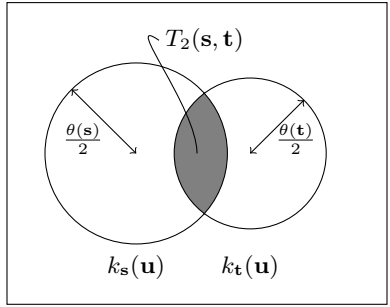
\begin{figure}
\begin{center}
\begin{tikzpicture}
\filldraw[fill=white] (-2,-2) rectangle (3,2);
\scope 
\clip (-0.3,0) circle (1.2);
\fill[gray] (1.2,0) circle (1);
\endscope
\draw (-0.3,0) circle (1.2) node [text=black] { }
      (1.2,0) circle (1) node [text=black] { };
\node at (0.6,1.5) {\footnotesize{$T_2(\mv{s},\mv{t})$}};
\node at (-0.3,-1.5) {\footnotesize{$k_\mv{s}(\mv{u})$}};
\node at (1.2,-1.5) {\footnotesize{$k_\mv{t}(\mv{u})$}};
\draw [<->] (-0.3,0) -- (-1.148528137423857, 0.848528137423857);
\draw [<->] (1.2,0) -- (1.907106781186548, 0.707106781186547);
\node at (-1,0.2) {\footnotesize{$\frac{\theta(\mv{s})}{2}$}};
\node at (1.8,0.2) {\footnotesize{$\frac{\theta(\mv{t})}{2}$}};
\draw (0.5,0) to [out=90,in=145] (0,1.5);
\end{tikzpicture}
\end{center}
\caption{The value of the non-stationary covariance function $T_2(\mv{s},\mv{t})$ is given by the area of the shaded region normalized by the areas of the two circles.}
\label{fig:T2}
\end{figure}

The perhaps simplest choice of $k_\mv{s}(\mv{u})$ is the indicator function
\begin{equation}\label{eq:kernel}
k_{2,\mv{s}}(\mv{u}) = \frac{1}{\theta(\mv{s})\sqrt{\pi}}\mathbb{I}\left(\|\mv{s}-\mv{u}\|<\frac{\theta(\mv{s})}{2}\right) = \begin{cases}
\frac{1}{\theta(\mv{s})\sqrt{\pi}} & \mbox{if } \|\mv{s}-\mv{u}\|<\theta(\mv{s})/2,\\
0 & \mbox{otherwise.}
\end{cases}
\end{equation}
The normalization of the kernel is chosen so that the resulting taper satisfies $T(\mv{s},\mv{s}) = 1$. In this parameterization, the taper range of a stationary model with $\theta(\mv{s})\equiv \theta$ is given by $\theta$. The value of $T(\mv{s},\mv{t})$ is in this case obtained as the normalized area of the asymmetric lens produced by the intersection of two circles centered at $\mv{s}$ and $\mv{t}$, with diameters $\theta(\mv{s})$ and $\theta(\mv{t})$ respectively. See Figure~\ref{fig:T2} for an illustration.

Straightforward calculations give that the resulting taper is
\begin{equation*}
T_2(\mv{s},\mv{t}) =
\frac1{\pi r_{st}R_{st}} \cdot \begin{cases}
\pi r_{st} & \mbox{$d_\mv{st} < R_{st} - r_{st}$}, \\
V_2\left(R_{st},\frac{d_{st}^2 + R_{st}^2 - r_{st}^2}{2d_{st}}\right) + V_2\left(r_{st},\frac{d_{st}^2 + r_{st}^2 - R_{st}^2}{2d_{st}}\right) & \mbox{$R_{st}-r_{st} \leq d_\mv{st} < R_{st} + r_{st}$},\\
0 & \mbox{otherwise}.
\end{cases}
\end{equation*}
Here $d_\mv{st} = \|\mv{s}-\mv{t}\|$, $R_{st} = \frac1{2}\max(\theta(\mv{s}),\theta(\mv{t}))$, and $r_{st} = \frac1{2}\min(\theta(\mv{s}),\theta(\mv{t}))$. Further,
\begin{equation}
V_2(r,x) = \begin{cases}
r^2\cos^{-1}\left(\frac{x}{r}\right) - x\sqrt{r^2-x^2} & |x| < r\\
0 & \mbox{otherwise,}
\end{cases}
\end{equation}
is the area of a circular cap with triangular height $x$ of a circle with radius $r$. The case $d_{\mv{st}} < R_{st} - r_{st}$ does not occur if the taper range $\theta(\mv{s})$ is Hölder continuous with exponent $1$. This can be a natural condition to require in applications where the taper range should vary smoothly across space.

With $\theta(\mv{s})\equiv\theta$, the taper function reduces to
\begin{equation*}
T(d) = \begin{cases}
\frac{2}{\pi}\cos^{-1}(\frac{d}{\theta}) - \frac2{\pi}\frac{d}{\theta}\sqrt{1 - \frac{d^2}{\theta^2}} & \mbox{for $d < \theta$},\\
0 & \mbox{otherwise},
\end{cases}
\end{equation*}
and it is easy to show that this is a valid covariance function for processes on $\R$ and on $\R^2$, but not for any $\R^m$ with $m>2$.

A variant of the taper is obtained if $\mv{s}$ and $\mv{u}$ are interpreted as vectors in $\R^3$. The normalized kernel is then
\begin{equation}\label{eq:kernel2}
k_{3,\mv{s}}(\mv{u}) = \frac{\sqrt{3}}{\sqrt{4\pi\theta(\mv{s})^3}}\mathbb{I}\left(\|\mv{s}-\mv{u}\|<\frac{\theta(\mv{s})}{2}\right),
\end{equation}
and the expression for the resulting taper is obtained as the normalized volume of the intersection of two spheres in $\R^3$,
\begin{equation*}
\resizebox{\linewidth}{!}{$
T_3(\mv{s},\mv{t}) =
\frac{3}{4\pi (r_{st}R_{st})^{3/2}} \cdot \begin{cases}
\frac{4\pi r_{st}^{3/2}}{3} & \mbox{$d_\mv{st} < R_{st} - r_{st}$}, \\
V_3\left(R_{st},\frac{d_{st}^2 + R_{st}^2 - r_{st}^2}{2d_{st}}\right) + V_3\left(r_{st},\frac{d_{st}^2 + r_{st}^2 - R_{st}^2}{2d_{st}}\right) & \mbox{$R_{st}-r_{st} \leq d_\mv{st} < R_{st} + r_{st}$},\\
0 & \mbox{otherwise}.
\end{cases}$}
\end{equation*}
Here, $V_3(r,x)$ denotes the volume of the spherical cap with triangular height $x$ of a sphere with radius $r$, which is given by
\begin{equation*}
V_3(r,x) =
\begin{cases}
\frac{\pi}{3}(r-x)^2(2r+x) & |x| < r,\\
0 & \mbox{otherwise}.
\end{cases}
\end{equation*}
With $\theta(\mv{s})\equiv \theta$, the taper simplifies to
\begin{equation*}
T(d) = \begin{cases}
\frac1{2\theta^3}(2\theta + d)(\theta - d)^2 & \mbox{if $d<\theta$,}\\
0 & \mbox{otherwise,}
\end{cases}
\end{equation*}
which is a valid covariance function on $\R^m$ for $m\leq 3$ and is commonly referred to as the spherical covariance function.

By the same reasoning on can create a taper function by defining the generating kernel as a normalized hypersphere in $\R^n$. The taper is then obtained as the normalized volume of the intersection of the two hyperspheres, which for a general $n$ is
\begin{equation*}
\resizebox{\linewidth}{!}{$
T_n(\mv{s},\mv{t}) =
\frac{\Gamma(n/2+1)}{(\pi r_{st}R_{st})^{n/2}} \cdot \begin{cases}
\frac{\pi^n r_{st}^{n/2}}{\Gamma(n/2+1)} & \mbox{if $d_\mv{st} < R_{st} - r_{st}$}, \\
V_n\left(R_{st},\frac{d_{st}^2 + R_{st}^2 - r_{st}^2}{2d_{st}}\right) + V_n\left(r_{st},\frac{d_{st}^2 + r_{st}^2 - R_{st}^2}{2d_{st}}\right) & \mbox{if $R_{st}-r_{st} \leq d_\mv{st} < R_{st} + r_{st}$}.\\
0 & \mbox{otherwise}.
\end{cases}$}
\end{equation*}
Here $\Gamma(x)$ is the gamma function and $V_n$ denotes the volume of a hyperspherical cap with triangular height $x$, of a hypersphere with radius $r$, which is given by
\begin{equation*}
V_n(r,x) = \frac{\pi^{n/2}r^n}{2\Gamma(n/2+1)} \cdot \begin{cases}
I_{1-(x/r)^2}\left(\frac{n+1}{2},\frac1{2}\right) & 0 \leq x < r,\\
1-I_{1-(x/r)^2}\left(\frac{n+1}{2},\frac1{2}\right) & -r < x < 0,\\
0 & \mbox{otherwise}.
\end{cases}
\end{equation*}
Here, $I_{x}$ is the regularized incomplete beta function, defined as
\begin{equation*}
I_x\left(\frac{n+1}{2},\frac1{2}\right) = \frac{\int_{0}^x \sqrt{t^{n-1}(1-t)^{-1}}dt}{\int_{0}^1 \sqrt{t^{n-1}(1-t)^{-1}}dt}.
\end{equation*}
See \cite{li2011concise} for a derivation of the volume of a hyperspherical cap.

\begin{figure}
\centerline{\includegraphics[width=0.6\textwidth]{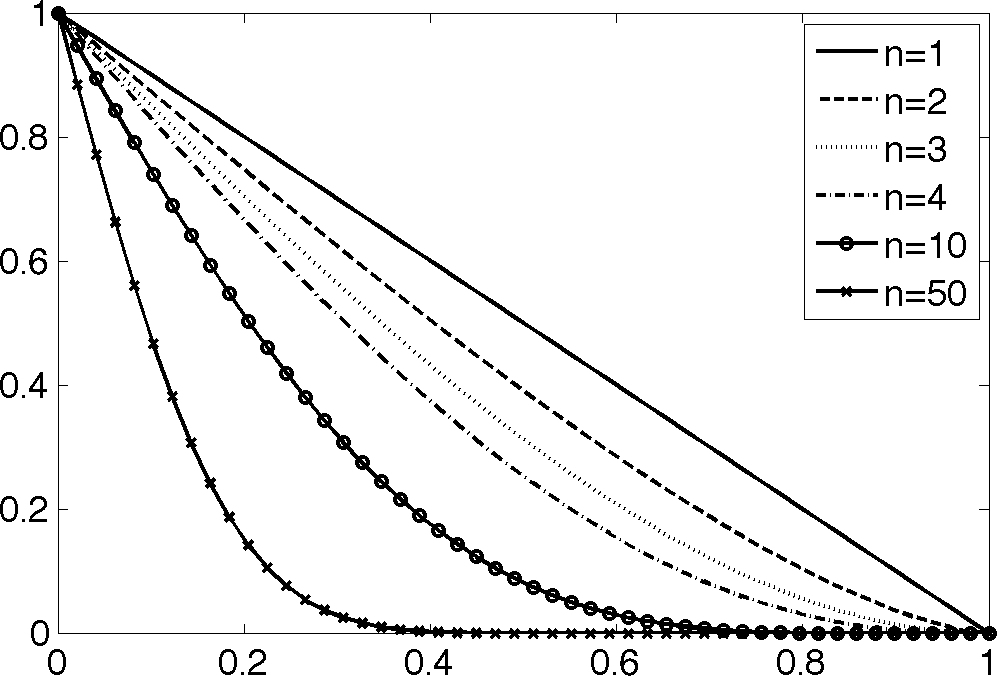}}
\caption{The stationary hyperspherical covariance functions for $\theta=1$ for different values of $n$.}
\label{fig:tapers}
\end{figure}

By construction, $T_n( \mv{s},\mv{t})$ is a valid covariance function for $\R^m$, $m\leq n$. Choosing $\theta(\mv{s}) \equiv \theta$ results in the stationary and isotropic correlation function
\begin{align}\label{eq:euclid}
T_n(d) = I_n\left(1-\left(\frac{d}{\theta}\right)^2\right),
\end{align}
which is shown for different values of $n$ in Figure \ref{fig:tapers}. This function is sometimes referred to as Euclid's hat and is frequently used in spatial statistics \citep[see e.g.][Chapter 3]{matern60}. The function is continuous at the origin and has $n-1$ derivatives at $d=\theta$ \citep{gneiting1999radial}. Furthermore,  Euclid's hat with $n=m$ is the solution to Tur\'an's problem, which means that it is the correlation function with the maximal integral over its support \citep[see][Section 4.4]{ehm2004}. This is important for tapering because one then wants to taper low values to zero while keeping large value relatively unaffected. Thus, these tapers are both simple to construct and have several desirable properties.

\subsection{Smooth tapers}
The differentiability at the origin of the taper function is an important property. When using a compactly supported covariance function for tapering, one generally wants it to has as many derivatives at the origin as the original covariance function \citep{furrer06}. A drawback with the family $T_n$ is therefore that all functions in the class have the same differentiability at the origin.

To obtain a smoother taper one can replace the spherical kernel with some differentiable function in \eqref{eq:kerneltaper}. For example, one could define an $m$ times differentiable kernel as the convolution of $m$ spherical kernels, $k_{n,\mv{s}}^m(\mv{u}) = k_{n,\mv{s}}(\mv{u}) \ast \ldots \ast k_{n,\mv{s}}(\mv{u})$. However, the problem with this construction, and most other convolution-based approaches, is that the integral in \eqref{eq:kerneltaper} in general is not available in closed form. We believe that one could compute the taper analytically if $k_{n,\mv{s}}^m(\mv{u})$ is used as a kernel, but it is not clear whether one can find some compact and simple expression for it.

An alternative is to do the integration numerically. However, as soon as numerical integration is required the computational cost for forming the taper matrix increases and can easily outweigh the cost of, for example, computing the kriging predictor using a full covariance matrix. We have not been able to find any non-stationary tapers based on isotropic kernels which are as easy to evaluate as the $T_n$ family, and will not pursue this issue further here. See for example \cite{mateu2013class} for other kernel-based approaches to constructing non-stationary compactly supported covariance functions.

A simpler alternative for constructing a more differentiable non-stationary covariance function is to drop the requirement that the kernel should be isotropic. One can then construct the kernel as a product $k_{\mv{s}}^m(\mv{u}) = \prod_{i=1}^d k_{s_i}^m(u_i)$, where $k_{s_i}^m(u) = k_{s_i}(u) \ast \ldots \ast k_{s_i}(u)$ is the convolution of $m$ spherical kernels on $\R$. The taper is then given by
\begin{equation}\label{eq:Tnk}
T^m(\mv{s}, \mv{t}) = \int k_{\mv{s}}^m(\mv{u}) k_{\mv{t}}^m(\mv{u}) d\mv{u} = \prod_{i = 1}^d \int k_{s_i}^m(u_i) k_{t_i}^m(u_i) du_i = \prod_{i = 1}^d T^m(s_i,t_i).
\end{equation}
The spherical kernel on $\R$ is simply an indicator function and $k_{s}^m$ is therefore a scaled and shifted cardinal B-spline. Thus, to evaluate $T^m(s_i,t_i)$ one has to compute the integral of the product of two B splines, which can be done using integration by parts \citep{vermeulen1992integrating}.

For $m=1$ one has $k_{s}^1(u) = \mathbb{I}(|u-s|<\theta(s)/2)$ and
\begin{equation*}
T^1(s, t) = \frac1{\sqrt{\theta(s)\theta(t)}}\begin{cases}
|s - t|, & \mbox{if $2|s-t|< \theta(s)+\theta(t)$}\\
0 & \mbox{otherwise.}
\end{cases}
\end{equation*}
This is the one-dimensional version of the Hyperspherical taper $T_1(s)$, and if $\theta(s) \equiv \theta$ it is the triangular covariance function. For $m=2$, the kernel instead is
\begin{equation*}
k_s^2(u) =  \frac{\sqrt{3}}{\sqrt{\theta(s)}}\begin{cases}
1 + 2\frac{u}{\theta(s)} & \mbox{if $s -\theta(s) < u < s$},\\
1 - 2\frac{u}{\theta(s)} & \mbox{if $s \leq u < s + \theta(s)$},\\
0 & \mbox{otherwise.}
\end{cases}
\end{equation*}
which gives the taper
\begin{equation*}
\resizebox{\linewidth}{!}{$
T^2(s,t) = \frac1{4(rR)^{3/2}}\cdot
\begin{cases}
2d^3 - 6d^2r + 2r^2(3R-r) & \mbox{$d \leq \min(r,R-r)$}, \\
3[d^3 - (d^2+rR)(R + r) + d(r-R)^2] -R^3 - r^3 & \mbox{$R-r<d<r$},\\
6r^2(R-d) & \mbox{$r<d\leq R-r$},\\
3[(rR-d^2)(r +R) + d(R - r)^2]  + r^3 - R^3 + d^3 & \mbox{$\max(r,R-r) < d \leq R$}, \\
3[(d^2+rR)(r + R) -d(r+R)^2] + R^3 + r^3 -d^3& \mbox{$R < d \leq R+r$},\\
0 & \mbox{otherwise.}
\end{cases}$}
\end{equation*}
Here, $d = |s-t|$, $r = \frac1{2}\min(\theta(s),\theta(t))$, and $R = \frac1{2}\max(\theta(s),\theta(t))$. If $\theta(s) \equiv \theta$, $T^2(s,t)$ simplifies to
\begin{equation*}
T^2(d) = \frac1{8\theta^3}\cdot
\begin{cases}
\theta^3 + 6d^3 -6\theta d^2 & \mbox{$0<d<\theta/2$},\\
2(\theta^3 - 3\theta^2 d + 3\theta d^2 - d^3) & \mbox{$\theta/2 < d \leq \theta$},\\
0 & \mbox{otherwise.}
\end{cases}
\end{equation*}

\section{Tapering and kriging prediction}\label{sec:kriging}
Let $X(\mv{s})$ be a mean-zero Gaussian random field with some covariance function $r(\mv{s},\mv{t})$. Suppose that $X$ is observed at the locations $\mv{s}_1, \ldots, \mv{s}_N$ and let $\mv{x}_0 = (X(\mv{s}_1), \ldots, X(\mv{s}_N))^{\trsp}$. Further, let $\mv{x}_1 = (X(\mv{s}_1^*), \ldots, X(\mv{s}_n^*))^\trsp$ be a vector with values that are to be predicted. The joint distribution of $\mv{x}_0$ and $\mv{x}_1$ can then be written as
\begin{equation*}
\begin{pmatrix}
\mv{x}_0\\
\mv{x}_1
\end{pmatrix}
 \sim
 \pN\left(\mv{0},
 \begin{pmatrix}
\mv{\Sigma}_{00} & \mv{\Sigma}_{01}\\
\mv{\Sigma}_{10} & \mv{\Sigma}_{11}
\end{pmatrix}
\right).
\end{equation*}
The conditional distribution of $\mv{x}_1$ given $\mv{x}_0$ is $\mv{x}_1 | \mv{x}_0 \sim \pN(\mv{\Sigma}_{10}\mv{\Sigma}_{00}^{-1}\mv{x}_0, \mv{\Sigma}_{11} - \mv{\Sigma}_{10}\mv{\Sigma}_{00}^{-1}\mv{\Sigma}_{01})$.
The mean $\hat{\mv{x}}_1 = \mv{\Sigma}_{10}\mv{\Sigma}_{00}^{-1}\mv{x}_0$ is the kriging predictor, and the MSE of the predictor is given by the diagonal of the covariance matrix, $\text{MSE}(\hat{\mv{x}}) = \diag(\mv{\Sigma}_{11} - \mv{\Sigma}_{10}\mv{\Sigma}_{00}^{-1}\mv{\Sigma}_{01})$.
As previously mentioned, solving $\mv{\Sigma}_{00}^{-1}\mv{x}_0$ is computationally expensive, and an alternative is to use a tapering estimate $\tilde{\mv{x}}_1 = \tilde{\mv{\Sigma}}_{10}\tilde{\mv{\Sigma}}_{00}^{-1}\mv{x}_0$, where the tilde indicates that the covariance matrices are based on the tapered covariance  $r(\mv{s},\mv{t})T(\mv{s},\mv{t})$. That is, $\tilde{\mv{\Sigma}}_{00} = \mv{\Sigma}_{00}\circ \mv{T}$, where $T_{ij} = T(\mv{s}_i, \mv{s}_j)$ and $\circ$ denotes the element-wise product. The MSE of the tapered prediction is given by
\begin{equation}\label{eq:mse}
\text{MSE}(\tilde{\mv{x}}) = \diag(\mv{\Sigma}_{11} - \tilde{\mv{\Sigma}}_{10}\tilde{\mv{\Sigma}}_{00}^{-1}\mv{\Sigma}_{01} - \mv{\Sigma}_{10}\tilde{\mv{\Sigma}}_{00}^{-1}\tilde{\mv{\Sigma}}_{01} + \tilde{\mv{\Sigma}}_{10}\tilde{\mv{\Sigma}}_{00}^{-1} \mv{\Sigma}_{00}\tilde{\mv{\Sigma}}_{00}^{-1}\tilde{\mv{\Sigma}}_{01}),
\end{equation}
which reduces to the MSE of the kriging predictor if  $T  (\mv{s},\mv{t}) = 1$. The MSE of the tapered prediction is a natural measure of accuracy which we will use to compare different methods in what follows.

\subsection{Using adaptive tapers}\label{sec:taperrange}
When tapering is used for kriging, the stationary taper range is usually set so that $\tilde{\mv{\Sigma}}_{00}$ has a fixed percentage of non-zero values. The goal with adaptive tapering is to achieve better predictions without increasing the number of non-zero values in $\tilde{\mv{\Sigma}}_{00}$. To do this, the local taper ranges $\theta(\mv{s})$ of the adaptive taper have to be chosen. The best choice of $\theta(\mv{s})$ would be a function that minimizes the kriging error while keeping the number of non-zero elements in $\tilde{\mv{\Sigma}}_{00}$ fixed. This is however difficult to do in practice, so we will instead split the problem in two parts. Since the sparsity of $\tilde{\mv{\Sigma}}_{00}$ only is affected by the values of $\theta(\mv{s})$ at the measurement locations $\mv{s}_1,\ldots, \mv{s}_N$, we start by choosing $\mv{\theta}_{{1:N}} = (\theta(\mv{s}_1), \theta(\mv{s}_2), \ldots, \theta(\mv{s}_N))$ in order to achieve a given sparsity. As a second step, we interpolate the values $\mv{\theta}_{{1:N}}$ to obtain the continuous function $\theta(\mv{s})$ which is used to calculate $\tilde{\mv{\Sigma}}_{10}$.
The details of these two steps are discussed in the following two subsections.

\subsubsection{Choosing the taper range at the measurement locations}
Ideally, $\mv{\theta}_{{1:N}}$ should be chosen so that every row in $\tilde{\mv{\Sigma}}_{00}$ has the same number, $m$, of non-zero elements. In other words, one would like the tapered covariance between $X(\mv{s}_i)$ and $X$ at the $m-1$ points nearest to $\mv{s}_i$ to be non-zero for each $i$. In general, one cannot get exactly $m$ elements for each row since the nearest-neighbor relation is not symmetric. However, $\mv{\theta}_{{1:N}}$ can be chosen so that the average number per row is $m$ while keeping the variation between the rows small. This can be achieved using Algorithm \ref{alg:thetaset}, where $n_j(\mv{\theta}_{{1:N}})$ is the number of non-zero elements of row $j$ in $\tilde{\mv{\Sigma}}_{00}$ for a given choice of taper ranges $\mv{\theta}_{{1:N}}$, and $r_j(\mv{\theta}_{1:N},k)$ denotes the $k$th largest value of the vector $(\|\mv{s}_j-\mv{s}_1\|,\|\mv{s}_j-\mv{s}_2\|,\ldots,\|\mv{s}_j-\mv{s}_N\|)^T-\mv{\theta}_{1:N}$. The parameter $\mv{m}_{1:N}= (m_1,\ldots,m_N)$ is a vector containing the desired number of non-zero elements for each row. The parameter $\epsilon$ determines the percentage of nodes that is allowed to have $|n_j(\mv{\theta}_{{1:N}})-m_j|>1$, and thus serves as a convergence criterium. Unless the measurement locations are structured in some malicious way, the algorithm will converge towards a point $\mv{\theta}_{1:N}$ that gives close to $m_j$ non-zero elements for each row $j$.

The algorithm works as follows: If $\tilde{\mv{\Sigma}}_{00}$ has too many non-zero elements at iteration $i$, the taper range of the location which corresponds to the row with the most non-zero elements is decreased so that that row gets the correct number of non-zero elements. If the matrix instead has too few non-zero elements at iteration $i$, the taper range of the location which corresponds to the row with the fewest number of non-zero elements is increased so that that row gets the correct number of non-zero elements. In the case that $\tilde{\mv{\Sigma}}_{00}$ has the correct number of non-zero elements, but the rows have different number of non-zero elements, the row with the largest deviation in the number of non-zero elements is changed. If several rows have the same deviation, one of these are picked at random at row $16$. In row $17$, random numbers are used since deterministic values can cause taper ranges for different points to be exactly equal, which can cause numerical problems. When the first loop has converged, the final loop goes through the nodes and maximizes the taper ranges while keeping the number of non-zero elements constant.

\begin{figure}
\begin{center}
\includegraphics[width=0.4\textwidth]{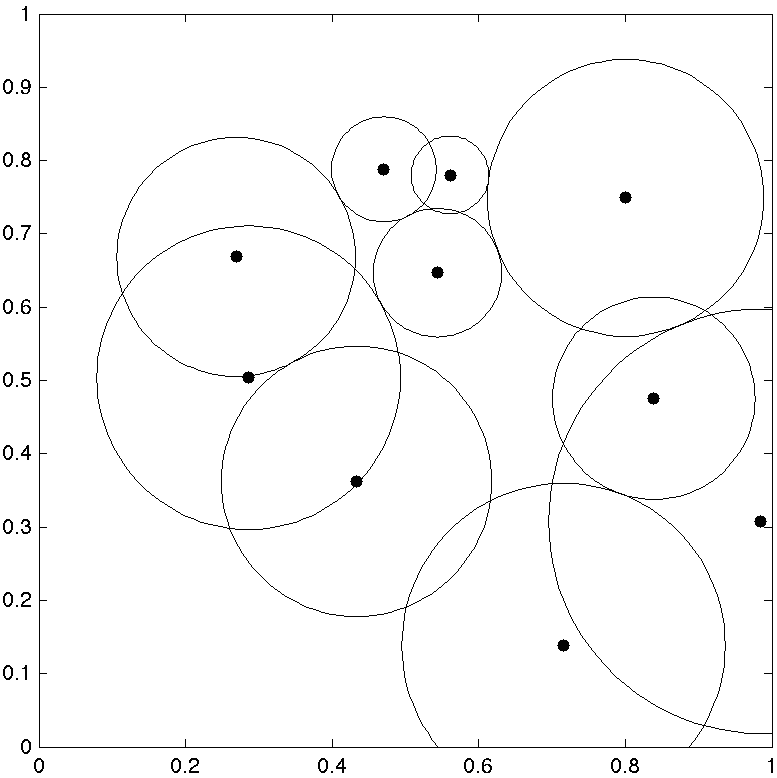}
\includegraphics[width=0.4\textwidth]{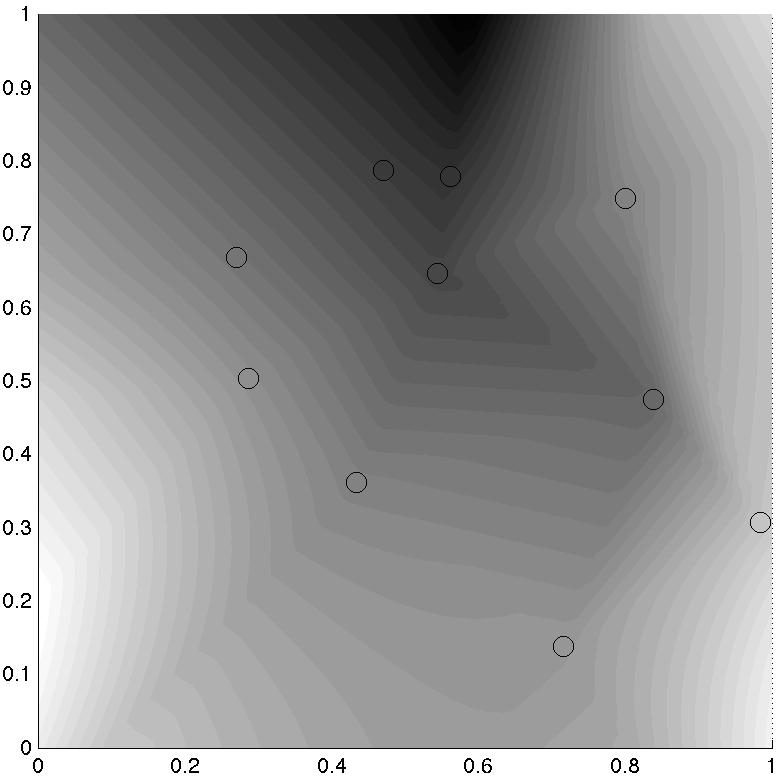}
\raisebox{0.4cm}{\includegraphics[width=0.05\textwidth]{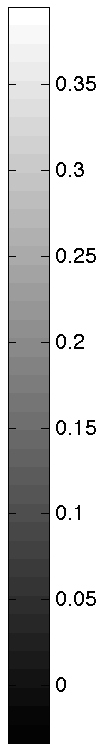}}
\end{center}
\caption{The left panel shows ten observation locations and the circular kernels $k_\mv{s}(\mv{u})$ for each observation location. The diameters $\theta(\mv{s})$ of the kernels are chosen using Algorithm \ref{alg:thetaset} so that each kernel overlaps with two other kernels, resulting in a tapering matrix with three non-zero elements per row. The right panel shows the taper range of the observation locations (the values in the circles) and an interpolated function which is used during kriging predictions of unobserved locations.}
\label{fig:taperrange}
\end{figure}

The result of the algorithm for a simple example with ten locations can be seen in the left panel of Figure \ref{fig:taperrange}. The result is visualized through the circular kernels $k_{\mv{s}}(\mv{u})$ for each location, where the radius of the circle at location $\mv{s}_i$ is given by $\theta_{\mv{s}_i}/2$.

 \begin{algorithm}[t]
 \caption{ }\label{alg:thetaset}
 \begin{algorithmic}[1]
 \Procedure{Theta-set}{$\mv{m}_{1:N}, \mv{\theta}_{{1:N}}^{(0)},  \mv{s}_{{1:N}}, \epsilon$}

  \State $i \gets 0$
    \Do
      \State $i \gets i+1$
      \State $\mv{\theta}_{{1:N}}^{(i)} \gets \mv{\theta}_{{1:N}}^{(i-1)}$
      \State $n_m \gets \min_l(n_l(\mv{\theta}_{{1:N}}^{(i)})-m_l)$
      \If{$\sum_l n_l(\mv{\theta}_{{1:N}}^{(j-1)}) = M $}
        \If{\scalebox{1}{$\sum_l |n_l(\mv{\theta}_{{1:N}}^{(i)})-m_l| =0  $ }}
          \State {\bf break}
        \ElsIf{$\max_l(n_l(\mv{\theta}_{{1:N}}^{(i)})-m_l) > |\min_l(n_l(\mv{\theta}_{{1:N}}^{(i)}) -m_l)|$}
            \State $n_m \gets \max_l(n_l(\mv{\theta}_{{1:N}}^{(i)})-m_l)$
        \EndIf
      \ElsIf{\scalebox{1}{$\sum_l n_l(\mv{\theta}_{{1:N}}^{(i)}) >Nm  $}}
        \State $n_m \gets \max_l(n_l(\mv{\theta}_{{1:N}}^{(i)})-m_l)$
      \EndIf
      \State $j \gets $ random integer from the set $\{l : n_l(\mv{\theta}_{{1:N}}^{(i)})-m_l= n_m\}$
      \State $\theta_{j}  \gets \proper{Unif} \left( r_j(\mv{\theta}_{{1:N}}^{(i)},[m_j]),r_j(\mv{\theta}_{{1:N}}^{(i)},[m_j]+1)\right)$
    \doWhile{$\{ \sum_{l} \mathbb{I}(|n_l(\mv{\theta}_{{1:N}}^{(i)}) - m_j| > 1) > N\epsilon \}$ }
    \For{$j=1,\dots,N$}
    \State $\theta_{j}  \gets r_j(\mv{\theta}_{{1:N}}^{(i)},n_j(\mv{\theta}_{{1:N}}^{(i)}))$
     \EndFor
      \State \textbf{return} $ \mv{\theta}_{{1:N}}^{(i)}$
 \EndProcedure
 \end{algorithmic}
 \end{algorithm}

\subsubsection{Interpolating the taper ranges}\label{sec:interp}
Running Algorithm \ref{alg:thetaset} results in a vector $\mv{\theta}_{{1:N}}$ that has to be interpolated in order to generate a continuous function $\theta(\mv{s})$. There are many ways to do this, and one of the simplest is to compute a Delaunay triangulation of the locations $\mv{s}_1,\ldots,\mv{s}_N$ and do linear interpolation within each triangle: For a location $\mv{s}$ within a triangle $\mathcal{T}$ with vertices $\mv{s}_a, \mv{s}_b, \mv{s}_c$ and corresponding taper ranges $\theta_a, \theta_b, \theta_c$, the interpolated taper range $\theta(\mv{s})$ is given by $\theta(\mv{s}) = w_a\theta_a +w_b\theta_b +w_c\theta_c$, where $\{w_a, w_b, w_c\}$ are the barycentric coordinates of $\mv{s}$ in $\mathcal{T}$. The result of this interpolation method, implemented using the \texttt{scatteredInterpolant} function in \cite{Matlab}, for a simple example with ten locations can be seen in the right panel of Figure \ref{fig:taperrange}.

One could imagine that more advanced interpolation methods, resulting in smoother surfaces, could improve the results when the method is used for kriging. However, initial studies indicate that the interpolation method has little effect on the resulting Kriging error. For example, the \texttt{scatteredInterpolant} function also supports a natural neighbor interpolation method that uses Voronoi tessellations to find a smoother interpolation ($C^1$ except at the measurement locations). In order to investigate if the kriging results could be improved by using more sophisticated interpolation methods, a part of the simulation study in Section \ref{sec:simulations} was done using both natural and linear interpolation. Specifically, the MSE of the kriging predictor using linear interpolation for one of the test cases using an exponential covariance function was $2.87$ (shown as $2.9$ in Table \ref{tab1}). Changing to the natural interpolation method instead resulted in an MSE of $2.78$. The mean time it took to compute the interpolation was $9.1ms$ for the linear interpolation and $12.6ms$ for the natural interpolation. Thus, it seems as if the results for the adaptive tapers could be improved slightly by using more complicated interpolation methods. However, since the difference is small, we use the simpler linear interpolation method for the rest of this work, and leave the question of how to optimally interpolate the taper ranges for future research.

\subsubsection{Examples using simulated data}\label{sec:examples}
\begin{figure}
\begin{center}
\includegraphics[width=0.4\textwidth]{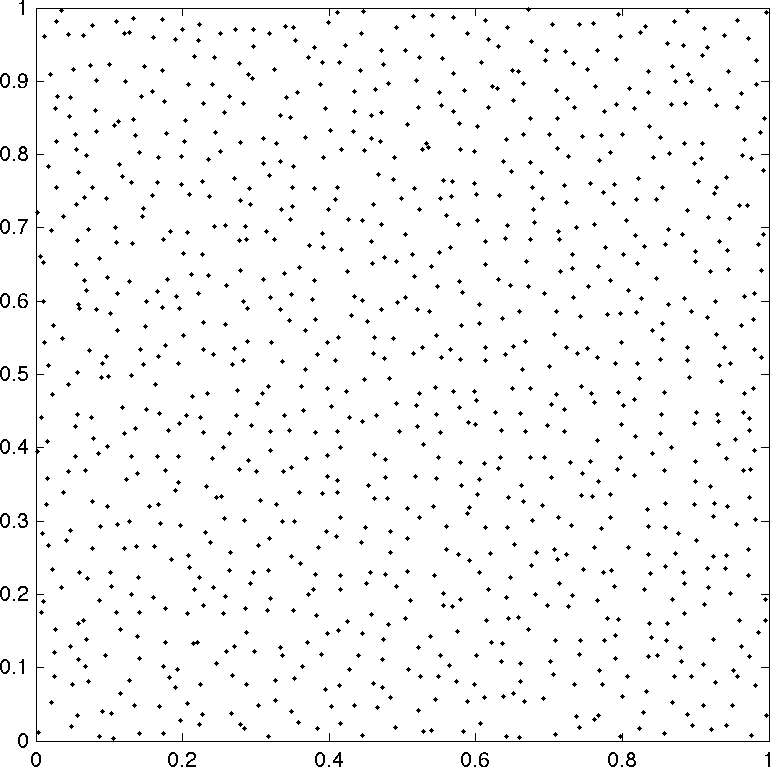}
\includegraphics[width=0.4\textwidth]{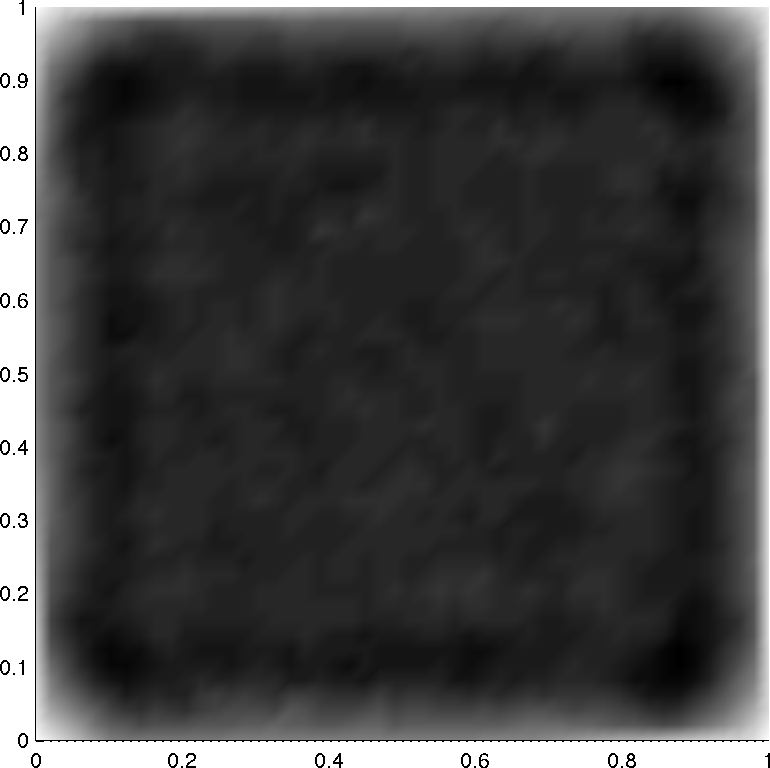}
\raisebox{0.3cm}{\includegraphics[width=0.045\textwidth]{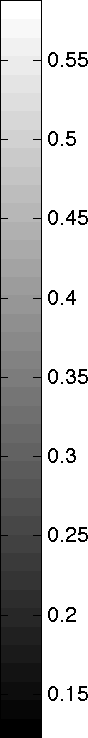}}
\\
\includegraphics[width=0.4\textwidth]{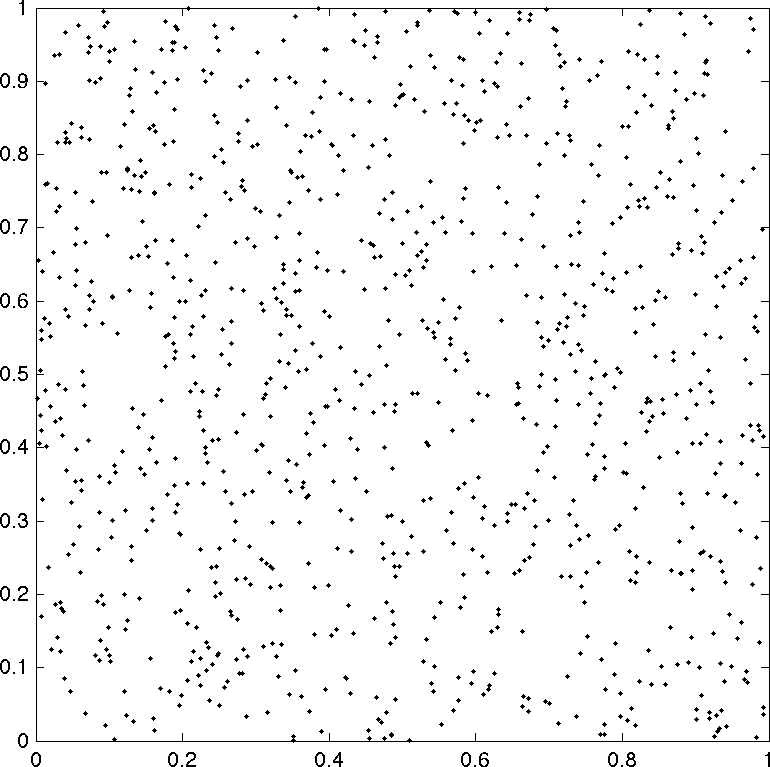}
\includegraphics[width=0.4\textwidth]{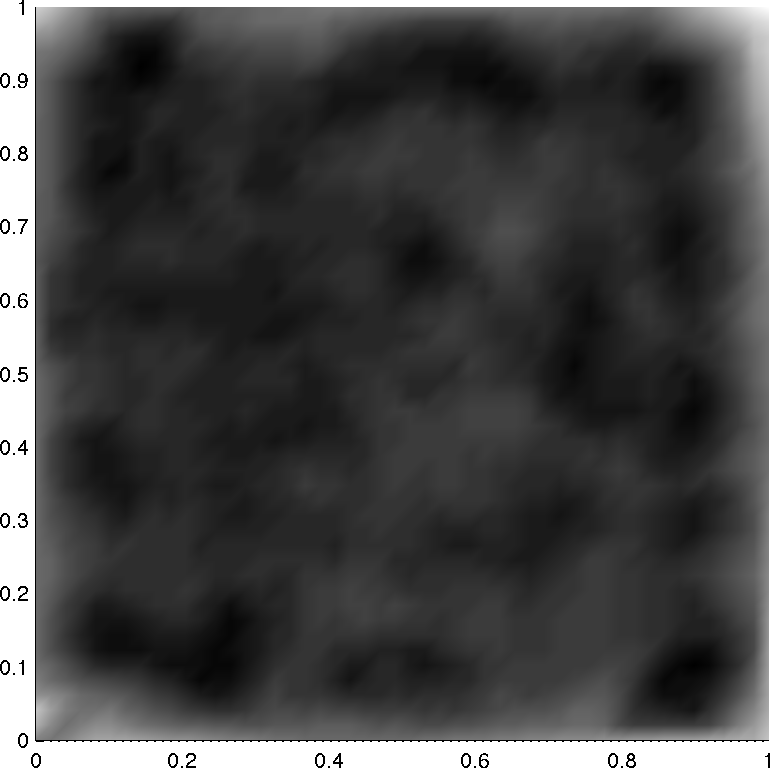}
\raisebox{0.3cm}{\includegraphics[width=0.045\textwidth]{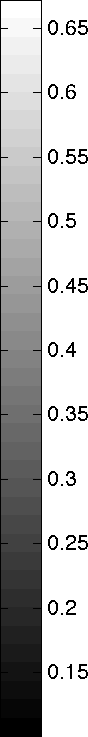}}
\\
\includegraphics[width=0.4\textwidth]{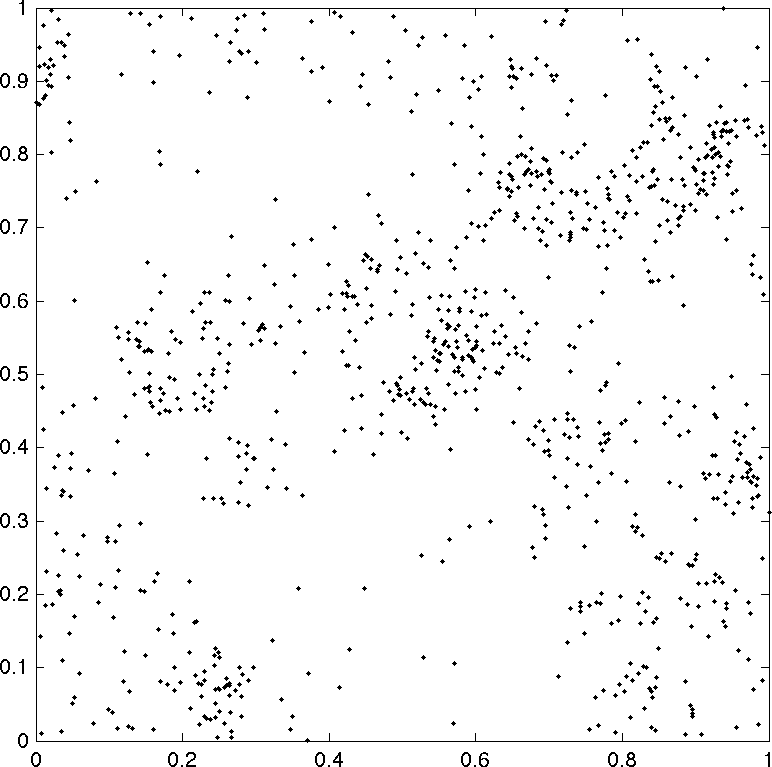}
\includegraphics[width=0.4\textwidth]{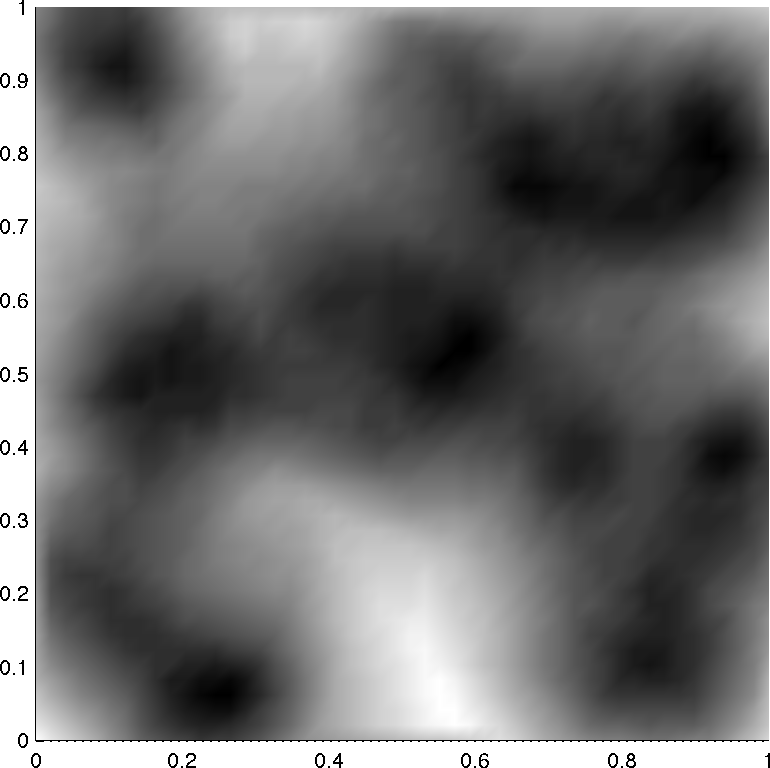}
\raisebox{0.3cm}{\includegraphics[width=0.038\textwidth]{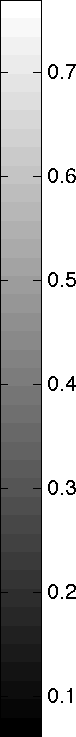}}
\vspace{-0.4cm}
\end{center}
\caption{The left panels show three types of measurement locations: Structured locations (top), completely spatially random locations (mid), and clustered locations (bottom). Each example has $1024$ measurement locations and the right panels show the corresponding non-stationary taper range for each scenario. The taper range is chosen using the method in Section \ref{sec:taperrange} so that the covariance matrix for the data has as many non-zero elements as if a  stationary taper range of $0.1$ was used.}
\label{fig:locations}
\end{figure}

Results of using Algorithm \ref{alg:thetaset} in combination with linear interpolation for three realistic scenarios are shown in Figure \ref{fig:locations}. The left panels show the measurement locations and the right panels show the corresponding taper ranges chosen so that the tapered covariance matrix has as many non-zero elements as if a stationary taper range of $0.1$ was used.

The top row shows a spatially structured case, where the measurements are taken at a perturbed grid. The measurements are done at the locations $\frac1{32}(0.5 + i + U_{ij}, 0.5 + j + V_{ij})$ for $i,j = 0, \ldots, 31$, where $U_{ij}$ and $V_{ij}$ are uniform random variables on $(-0.45, 0.45)$. Thus, this scenario is similar to that used in \cite{stein2013taper}. For this scenario, the taper range is larger at the edges of the domain but fairly constant in the interior, and the results are fairly similar to the warping method proposed by \cite{stein2013taper}. The second scenario, shown in the middle row of Figure \ref{fig:locations}, is complete spatial randomness, where the measurements are taken at $(U_{ij}, V_{ij})$ where $U_{ij}$ and $V_{ij}$ now are uniform random variables on $(0,1)$. Finally, the third scenario shown in the bottom row in the figure, corresponds to clustered locations, where the measurements locations are drawn from a log-Gaussian cox process. That is, the locations are drawn from a Poisson process with intensity $\lambda(\mv{s}) = \exp(Z(\mv{s}))$ where $Z(\mv{s})$ is a mean-zero Gaussian Mat\'ern field, with covariance function
\begin{equation}\label{eq:matern}
C(\mv{h}) = \frac{\sigma^2 2^{1-\nu}}{\Gamma(\nu)}(\kappa\|\mv{h}\|)^{\nu}K_{\nu}(\kappa\|\mv{h}\|), \quad \mv{h} \in \R^d, \nu>0.
\end{equation}
Here $\kappa$ is a range parameter, $\sigma^2$ the variance, $\nu$ is a smoothness parameter, and $K_{\nu}$ is a modified Bessel function. The distance where the covariance is approximately one tenth of the variance, sometimes referred to as the practical range, of the field is given by $\rho = \sqrt{8\nu}\kappa^{-1}$. For the scenario in Figure \ref{fig:locations}, the parameter $\kappa = 10$, $\nu = 0.5$, and $\sigma = 2$ were used.

Compared with the spatially structured case one can note larger variations in the taper ranges in the two other scenarios. The warping method in \cite{stein2013taper} cannot be used to construct tapers similar to our adaptive tapers in these cases, but one would then have to use the more complicated and computationally expensive warping methods by \cite{anderes2013}.

\begin{figure}
\begin{center}
\includegraphics[width=0.32\textwidth]{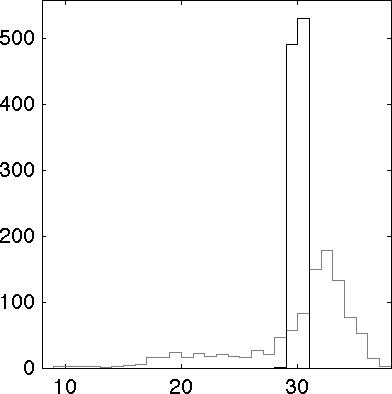}
\includegraphics[width=0.32\textwidth]{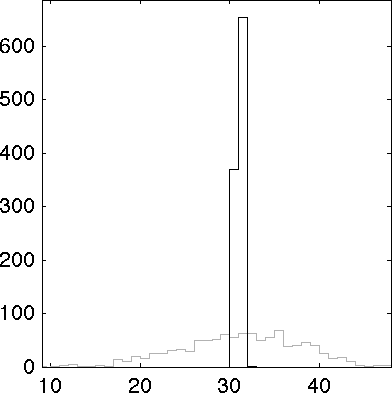}
\includegraphics[width=0.32\textwidth]{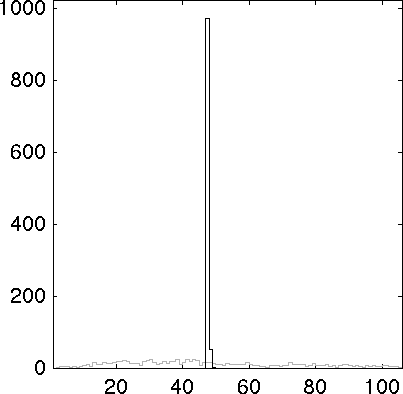}
\end{center}
\caption{Histograms of the number of non-zero elements per row in the tapered covariance matrix for data for the three examples in Figure \ref{fig:locations}: Structured locations (left), random locations (middle), and clustered locations (right). The grey curves show the stationary tapers and the black curves show the adaptive tapers.}
\label{fig:histograms}
\end{figure}

Histograms of the number of non-zero elements per row can be seen in Figure \ref{fig:histograms}. The figure has one panel for each of the scenarios in Figure \ref{fig:locations}, and each panel shows the histogram for the adaptive taper in black and the histogram for the stationary taper in grey. The total number of non-zero elements is the same for the two methods, but the variability of the number of non-zero elements for each row is much smaller for the adaptive tapers, especially in the clustered scenario. The variability of the number of non-zero elements per row for the adaptive taper is mainly caused by the fact that the average number of non-zero elements per row is not an integer. For example, for the leftmost panel it was approximately $29.6$, and in order to get the correct average the method needs to let the rows have different number of non-zero elements.

\subsection{Numerical comparisons of kriging prediction}\label{sec:simulations}
In this section, stationary and adaptive tapering approximations are compared in the case of kriging prediction for a Gaussian Mat\'ern field on $[0,1]^2$ with known parameters. The field is observed at $N$ locations in the domain and the value of the field is predicted for all locations on a $50 \times 50$ regular lattice in the domain. The comparisons are done both with $\nu=0.5$, which corresponds to an exponential covariance function, and with $\nu=1.5$ which corresponds to a smoother covariance function. For both values of $\nu$, one case with a long correlation range, $\rho = 0.2$, and one case with a shorter range, $\rho = 0.1$, are tested. For all cases, $\sigma = 1$ is used. For each parameter configuration, the three different measurement scenarios described in Section \ref{sec:examples} are tested, each with $N=1024$ measurement locations.

Five different tapering functions are compared. The first is a stationary Wendland function
\begin{equation*}
r_\theta(h) = \left(1-\frac{|h|}{\theta}\right)^4\left(1+4\frac{|h|}{\theta}\right)\mathbb{I}(|h|<\theta),
\end{equation*}
the second is a stationary hyperspherical taper \eqref{eq:euclid} with $n=2$, and the third is the corresponding non-stationary hyperspherical taper $T_2$. The final two are the non-stationary product tapers $T^1$ and $T^2$, where the latter has the same differentiability at the origin as the Wendland function. In order to get comparable results, the various taper ranges have to be chosen so that the methods have similar computational costs. To do this, the taper ranges of the two stationary tapers are set to $0.1$. Algorithm \ref{alg:thetaset} is then used to choose the non-stationary taper ranges so that all tapered covariance matrices have the same number of non-zero elements.

In summary, four different parameter settings and three different measurement scenarios are tested. This gives twelve test cases in total on which all five tapering methods are tested. For each test case, $100$ different data sets are simulated, and the optimal kriging predictor and the different tapered approximations are computed for each data set.

The relative increase in the MSE, compared to the optimal kriging predictor, is used as a measure of accuracy
\begin{equation}\label{eq:mse1}
\frac1{100n}\sum_{j=1}^{100}\sum_{i=1}^{n}\frac{\text{MSE}(\tilde{\mv{x}}^{(j)}_0(\mv{s}_i)) -\text{MSE}(\hat{\mv{x}}^{(j)}_0(\mv{s}_i))}{\text{MSE}(\hat{\mv{x}}^{(j)}_0(\mv{s}_i))},
\end{equation}
where $\tilde{\mv{x}}^{(j)}_0(\mv{s}_i)$ is the tapered kriging prediction at location $\mv{s}_i$ for the simulated dataset $j$. These values are shown in Table \ref{tab1} for the exponential covariance function, and in Table \ref{tab2} for the Mat\'ern covariance function.

\begin{table}[t]
\centering
\begin{tabular}{@{}lcccccc@{}}
\toprule
& \multicolumn{3}{c}{Short range} & \multicolumn{3}{c}{Long range} \\
& S & R & C & S & R & C \\
\cmidrule(r){2-4} \cmidrule(r){5-7}
$W(h)$ & $3.1$ ($0.06$) & $5.9$ ($0.18$) & $9.3$ ($0.42$) & $6.3$  ($0.11$) & $12.3$ ($0.52$) & $26.6$ ($2.15)$ \\
$T_2(h)$ & $1.2$ ($0.02$) & $2.3$ ($0.08$) & $4.7$ ($0.34$) & $1.8$  ($0.05$) & $3.7$ ($0.22$) & $12.5$ ($1.64)$ \\
$T_2(s,t)$ & $\mv{1.1}$ ($0.02$) & $\mv{2.0}$ ($0.08$) & $\mv{1.2}$ ($0.33$) & $\mv{1.5}$  ($0.03$) & $\mv{2.9}$ ($0.15$) & $\mv{1.8}$ ($0.58)$ \\
$T^1(s,t)$ & $1.6$ ($0.02$) & $2.5$ ($0.09$) & $1.5$ ($0.37$) & $2.8$  ($0.04$) & $4.0$ ($0.15$) & $2.6$ ($0.65)$ \\
$T^2(s,t)$ & $3.3$ ($0.06$) & $5.7$ ($0.22$) & $2.6$ ($0.88$) & $7.3$  ($0.11$) & $12.2$ ($0.50$) & $5.8$ ($2.12)$ \\
\bottomrule
\end{tabular}
\caption{The relative increase in MSE shown in percent (equation \eqref{eq:mse1} scaled by a factor $100$), for the different tapered kriging predictors compared to the optimal kriging predictor for the exponential covariance function. The results are shown for the Wendland taper (W), the stationary hyperspherical taper $T_2(h)$, the non-stationary hyperspherical taper $T_2(s,t)$, and the two product tapers $T^1(s,t)$ and $T^2(s,t)$ for structured (S), random (R), and clustered (C) sampling scenarios. The values in the parentheses are the Monte Carlo standard deviation for each estimate. The bold values indicate the best method for each case. }
\label{tab1}
\end{table}

\begin{table}[t]
\centering
\begin{tabular}{@{}lcccccc@{}}
\toprule
& \multicolumn{3}{c}{Short range} & \multicolumn{3}{c}{Long range} \\
& S & R & C & S & R & C \\
\cmidrule(r){2-4} \cmidrule(r){5-7}
$W(h)$ & $6.7$ ($0.22$) & $16.0$ ($0.58$) & $23.7$ ($1.23$) & $\mv{22.5}$  ($0.90$) & $60.5$ ($3.43$) & $131.7$ ($8.25)$ \\
$T_2(h)$ & $14.8$ ($0.20$) & $21.6$ ($0.50$) & $23.7$ ($0.81$) & $46.7$  ($0.60$) & $62.4$ ($1.52$) & $89.9$ ($4.28)$ \\
$T_2(s,t)$ & $15.0$ ($0.23$) & $21.9$ ($0.56$) & $14.7$ ($2.65$) & $45.1$  ($0.56$) & $\mv{59.9}$ ($1.62$) & $42.3$ ($6.59)$ \\
$T^1(s,t)$ & $20.0$ ($0.26$) & $26.2$ ($0.58$) & $17.5$ ($2.96$) & $61.4$  ($0.73$) & $73.6$ ($1.81$) & $52.1$ ($7.12)$ \\
$T^2(s,t)$ & $\mv{6.5}$ ($0.24$) & $\mv{15.3}$ ($0.82$) & $\mv{6.3}$ ($2.34$) & $26.8$  ($1.18$) & $64.1$ ($4.31$) & $\mv{28.3}$ ($10.61)$ \\
\bottomrule
\end{tabular}
\caption{The relative increase in MSE shown in percent (equation \eqref{eq:mse1} scaled by a factor $100$), for the different tapered kriging predictors compared to the optimal kriging predictor for the Mat\'ern covariance function.  The bold values indicate the best method for each case. See Table \ref{tab1} for explanations of the different test cases.}
\label{tab2}
\end{table}

There are several things to note in the tables. For the exponential covariance the non-stationary Hyperspherical taper is always better than the stationary tapers. The improvement is relatively small for structured observation locations, where is mainly comes from better predictions near the boundary of the observation domain, but the improvement is larger for spatially random locations and the largest for clustered observation locations. One should generally choose a tapering function which has the same differentiability as the true covariance function \citep[see e.g.][]{furrer06}, and this is the reason for why the hyperspherical tapers outperform the Wendland taper and the product taper $T^2$ in the exponential case. Also for the Mat\'ern covariance, the adaptive tapers perform the best in general. In this case, however, the product taper $T^2$ outperforms the hyperspherical taper. The reason for this is that the hyperspherical tapers do not satisfy the tapering condition by \cite{furrer06} for this case, and they therefore do not have asymptotically equivalent MSEs to the optimal kriging predictor. However, it is interesting to note that the non-stationary hyperspherical taper outperforms the Wendland taper in the clustered situations, despite the fact that it does not satisfy the tapering condition. It is also interesting to note that $T^2$ is best overall for the Mat\'ern case, despite the fact that it is anisotropic. Thus, non-stationarity of the tapers seems to be important for the accuracy of the predictions.

Table \ref{tab3} shows timing results of the different methods in the simulation study, implemented in \cite{Matlab} and performed using a Macbook Pro computer with a 2.6GHz Intel Core i7 processor (Apple Inc., Cupertino, CA, USA). The average time for computing $\theta(\mv{s})$ was $228ms$, of which approximately $219ms$ was spent on running Algorithm \ref{alg:thetaset}, and the rest was used for interpolating the result. However, the timing for this step is dependent on which measurement scenario that is used, and the structured scenario resulted in the fastest average computation time ($128ms$) while the clustered was the slowest ($376ms$). The time for constructing the covariance matrices, given the taper range, was very similar for the different methods, and comparable to the time it took to run Algorithm \ref{alg:thetaset}. Also the time it took to compute the kriging predictor was similar for the different methods, and the reason for this is that all tapering matrices had the same number of non-zero elements. It should be stressed here that the timings for the first two steps are highly implementation dependent, and especially the time it takes to run Algorithm 1 could be greatly reduced by implementing the procedure in C or Fortran.

\begin{table}[t]
\centering
\begin{tabular}{@{}lccc@{}}
\toprule
& Compute $\theta(\mv{s})$ & Construct $\mv{T}$ & Kriging\\
$W(h)$     & $-$   & $445$ & $32$   \\
$T_2(h)$   & $-$   & $741$ & $8$    \\
$T_2(s,t)$ & $228$ & $797$ & $11$ \\
$T^1(s,t)$ & $228$ & $642$ & $28$  \\
$T^2(s,t)$ & $228$ & $817$ & $27$  \\
\bottomrule
\end{tabular}
\caption{Average timings in milliseconds for the different steps in the simulation study. The first column shows the time for computing the non-stationary taper range using Algorithm 1 and interpolation. The second column shows the time for constructing the tapering matrix, and the third column shows the time for computing the kriging predictor using the tapered covariance matrix.}
\label{tab3}
\end{table}

\section{Tapering and parameter estimation}\label{sec:estimation}
In the previous section, tapering was used for spatial prediction based on measurements of a Gaussian random field with known covariance function. In practice the parameters, $\mv{\Psi}$, of the covariance function often have to be estimated from the data first, and we now look at how tapering can be used in this step. As before, $\mv{x}_0$ denotes a vector of observations of a mean-zero Gaussian random field $X(\mv{s})$. The log-likelihood, that should be maximized to estimate the parameters, is $l(\mv{\Psi}; \mv{x}_0) = -\frac1{2}\log|\mv{\Sigma}_{00}| - \frac1{2}\mv{x}_0^\trsp \mv{\Sigma}_{00}^{-1}\mv{x}_0$. The computational cost for evaluating the likelihood is $\mathcal{O}(N^{3})$ due to both the log-determinant and the matrix inverse. The natural tapering approximation of the likelihood is to replace $\mv{\Sigma}_{00}$ with the tapered covariance matrix $\tilde{\mv{\Sigma}}_{00}$. However, this can cause substantial bias in the resulting estimates \citep{kaufman2008}, so a common choice is to instead use the tapered likelihood $\tilde{l}(\mv{\Psi}; \mv{x}_0) = -\frac1{2}\log|\tilde{\mv{\Sigma}}_{00}| - \frac1{2}\mv{x}_0^\trsp(\tilde{\mv{\Sigma}}_{00}^{-1}\circ \mv{T})\mv{x}_0$.

In the case of kriging estimation, adaptive tapers with taper ranges that adapted to the measurement locations was used to improve the predictions. For parameter estimation, however, it is not clear whether this is the best option.
We therefore propose the following method for selecting tapering ranges in order to generate a taper matrix with $M$ non-zero elements. The method, denoted $\alpha$-adaptive tapering, can be used as a tradeoff between fully adaptive tapering and stationary tapering.

Let $\mv{m}_s = (m_1, \ldots, m_N)$ be a vector where $m_i$ is the number of non-zero elements for the $i$th row in $\mv{T}$ for a stationary taper with taper range chosen so that $\mv{T}$ has $M$ non-zero elements in total, and select $\alpha \in [0,1]$. Then, compute the adaptive taper ranges by running Algorithm \ref{alg:thetaset} with $\mv{m} = (1-\alpha)\mv{m}_s + \alpha M/N$ as the vector of desired elements per row. Using $\alpha=0$ results in a stationary taper whereas $\alpha = 1$ results in the fully adaptive taper. Note that the method does not depend on the covariance matrix that is to be tapered, which means that the tapering matrix can be computed once before estimating the parameters.

\subsection{Numerical comparisons for parameter estimation}
In this section, adaptive tapering is tested in the case of parameter estimation for two exponential covariance models. Since exponential covariances are used, the hyperspherical taper $T_2$ is used in combination with the $\alpha$-adaptive method for selecting the taper ranges. As a reference method, the observations are partitioned into blocks and the true log-likelihood is approximated by summing the log-likelihoods of the blocks. This can be implemented as a block-diagonal taper with elements $T_{ij} = 0$ if $\mv{s}_i$ and $\mv{s}_j$ are in the same block, and $T_{ij} = 0$ otherwise. Because of this, \cite{stein2013taper} refers to the method as block tapering.

The first test case is estimation of a stationary exponential covariance model,  \eqref{eq:matern} with parameters $\nu=0.5$, $\kappa = 10$, and $\sigma = 10$. The second test case is a non-stationary exponential covariance model. Specifically, the approach by \cite{paciorek2006spatial} is used to produce a non-stationary exponential covariance function
\begin{equation}\label{eq:maternns}
C(\mv{s}_i,\mv{s}_j) = \sigma^2 |\mv{\Sigma}(\mv{s}_j)|^{1/4}|\mv{\Sigma}(\mv{s}_j)|^{1/4}\left|\frac{\mv{\Sigma}(\mv{s}_j)+\mv{\Sigma}(\mv{s}_j)}{2}\right|^{-1/2} \exp \left(-\sqrt{ Q_{ij}}\right).
\end{equation}
Here $\sigma^2$ is the variance, which is set to $4$ in the comparison, and
\begin{equation*}
Q_{ij} = (\mv{s}_i - \mv{s}_j)^T\left(\frac{\mv{\Sigma}(\mv{s}_j)+\mv{\Sigma}(\mv{s}_j)}{2}\right)^{-1}(\mv{s}_i - \mv{s}_j).
\end{equation*}
For simplicity, we choose $\mv{\Sigma}(\mv{s}) = \kappa(s)\mv{I}$ where $\kappa(s)$ is the spatially varying correlation range, modeled using a basis expansion $\log\kappa(\mv{s}) = \log\kappa(s_x,s_y) = \kappa_1 + \kappa_2 s_x$ where $\kappa_1 = -6$ and $\kappa_2 = 6$ are used. This results in a model where the covariance range is constant in the $s_y$ direction but varies in the $s_x$ direction.

For both the stationary and non-stationary test cases, data is generated by sampling a mean-zero Gaussian process $X(\mv{s})$ with the respective covariance function at $N=1024$ locations on $[0,1]^2$, chosen at random according to the spatially structured case used in Section \ref{sec:simulations}. The number of blocks for the block taper and the taper ranges for the $\alpha$-adaptive method are chosen so that all taper matrices have approximately $0.01N^2$ non-zero elements.

\begin{figure}
\begin{center}
\includegraphics[width=0.49\textwidth]{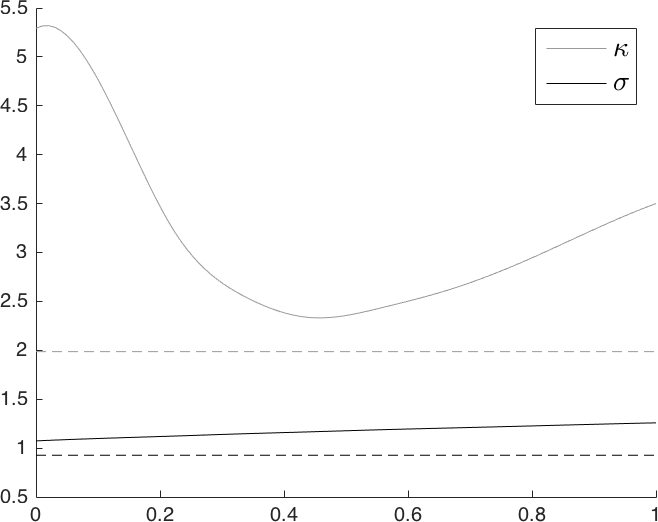}
\includegraphics[width=0.49\textwidth]{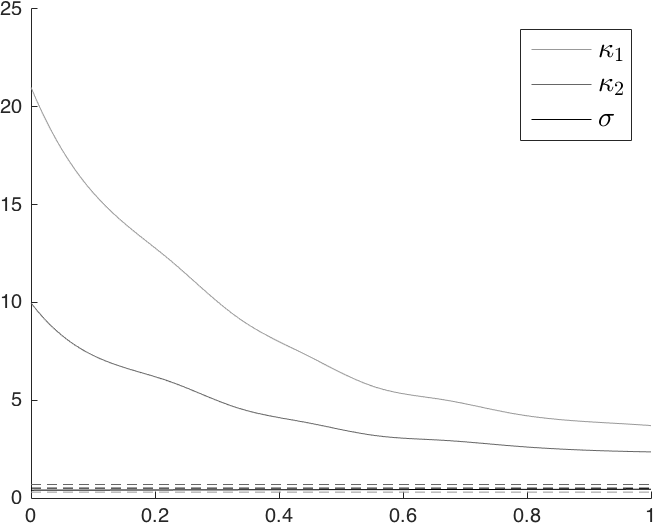}
\end{center}
\caption{Tapering results for the stationary (left panel) and non-statioary (right panel) exponential models. Square roots of the diagonal elements of the inverse Godabme information matrix for the $\alpha$-adaptive taper as a function of $\alpha$ (solid lines), and for the block taper (dashed lines). Note that ordinary stationary tapering corresponds to the case $\alpha = 0$.}
\label{fig:est1}
\end{figure}

We use the diagonal of the inverse Godambe information matrix \citep[see e.g.][]{stein2013taper} as a measure of the efficiency of the estimation methods based on different tapering functions. The elements correspond to the asymptotic marginal variances of the estimates obtained using the tapered likelihoods.
The Godambe information for the tapered loglikelihood is $\mv{G}(\mv{ \Psi}) = \pE_\mv{ \Psi}(\mv{H})\pE_\mv{ \Psi}(\mv{v}\mv{v}^{\trsp})^{-1}\pE_\mv{ \Psi}(\mv{H})$, where $\mv{v} = \nabla_\mv{ \Psi} \tilde{l}(\mv{\Psi}; \mv{x}_0)$ and $\mv{H}= \Delta_\mv{ \Psi} \tilde{l}(\mv{\Psi}; \mv{x}_0)$. The two expectations can be computed as
$\pE_\mv{ \Psi} (\mv{v}\mv{v}^{\trsp})_{ij} = \frac1{2}\trace\left((\mv\Sigma_i^{-1}\circ\mv T)\mv\Sigma_{00}(\mv\Sigma_j^{-1}\circ\mv T)\mv\Sigma_{00} \right)$
and $\pE_\mv{ \Psi} (\mv{H})_{ij} = \frac1{2}\trace\left(\mv{\Sigma}_i^{-1}\mv{\Sigma}_j\right)$, where $\mv{\Sigma}_i^{-1} = - \tilde{\mv{\Sigma}}_{00}^{-1}\mv\Sigma_i \tilde{\mv{\Sigma}}_{00}^{ -1}$ and
$
\mv{\Sigma}_i = \frac{\pd \tilde{\mv{\Sigma}}_{00}}{\pd \Psi_i} = \frac{\pd \mv\Sigma_{00}}{\pd  \Psi_i}\circ \mv{T}.
$

Figure \ref{fig:est1} displays the square roots of the diagonal elements of the inverse Godambe information matrix, as functions of $\alpha$ for the $\alpha$-adaptive method together with the corresponding values for the block taper for the two test cases. Recall that $\alpha = 0$ corresponds to stationary tapering whereas $\alpha = 1$ corresponds to fully adaptive tapering that minimizes the variation of non-zero elements between rows in the tapering matrix. The left panel of the figure shows the test case for the stationary exponential model, and one can note that the adaptive tapers outperform the stationary taper, and that an optimal value of $\alpha$ in this case is around $0.4$. The right panel shows the non-stationary test case. The adaptive tapers outperform the stationary taper also in this test case, and the optimal value is in this case $1$, meaning that the fully adaptive taper should be used. This is of little value, however, as the block taper is even better than the $\alpha$-adaptive tapers for both cases.

These are of course just two examples and different choices of true covariance function and measurement locations would have resulted in different optimal values for $\alpha$. However, the general conclusion has been the same for most cases we have tried so far: Non-stationary tapering outperforms stationary tapering, but the block tapers outperform both.

\section{Discussion}\label{sec:conclusions}
We have presented a class of computationally convenient compactly supported non-stationary covariance functions. Together with the proposed methods for selecting non-stationary taper ranges, these functions can be used for  spatially adaptive covariance tapering.

Numerical experiments showed that the adaptive tapering method can improve the performance of tapering for both kriging and parameter estimation, with negligible increase in computational cost. For parameter estimation, however, the numerical experiments also showed that dividing the data into blocks and ignoring the dependence between the blocks is often a better method than tapering. Thus, even if adaptive tapering is better than stationary tapering, we agree with \cite{stein2013taper} that tapering seems to have little practical value for parameter estimation.

An alternative to using the adaptive tapers for parameter estimation in applications is to use them directly as non-stationary covariance models for the data. In this case, an interesting model extension is to allow for anisotropy in the covariance function by using elliptical kernels in the construction. With this straightforward extension, the model could be used as a computationally efficient alternative to the non-stationary models by \cite{paciorek2006spatial}. Another relevant extension is to derive the expressions for the differentiable tapers, obtained using kernels that are convolutions of the spherical kernels. This would further increase the value of this class of covariance models. Finally, another interesting direction of future research is to use the adaptive tapers in an adaptive MCMC method similar to the method by \cite{wallin15adaptive}.


\section*{Acknowledgements}
The authors are grateful to the reviewers for their helpful and
constructive comments on earlier versions of the manuscript. Both authors have been supported by the Knut and Alice Wallenberg foundation and the second author has been supported by the Swedish Research Council Grant 2008-5382.

\bibliographystyle{elsarticle-harv}
\bibliography{journ_abrv,completebib}
\end{document}